\documentclass[12pt,a4paper]{article}

\usepackage{moreverb,dblfloatfix}
\usepackage{amssymb}
\usepackage{verbatim}
\usepackage[fleqn]{amsmath}
\usepackage{graphicx}
\usepackage{subfigure}
\usepackage{color}
\usepackage{float}
\usepackage{algorithmic}
\usepackage[linesnumbered,ruled]{algorithm2e}
\usepackage{amsmath}
\usepackage{fullpage}
\usepackage{upgreek}
\usepackage[numbers]{natbib}  
\usepackage{xspace}
\usepackage{listings}
\usepackage{authblk}

\newcommand{\x}{\mathbf{x}}
\newcommand{\xa}{\mathbf{x}^\textnormal{a}}
\newcommand{\xb}{\mathbf{x}^\textnormal{b}}

\newcommand{\y}{\mathbf{y}}

\newcommand{\nobs}{\textsc{n}_\textnormal{obs}}
\newcommand{\Nobs}{\textnormal{m}}
\newcommand{\Nens}{\textsc{n}_\textnormal{ens}}
\newcommand{\nvar}{\textsc{n}_\textnormal{var}}
\newcommand{\nc}{\textsc{n}_\textnormal{c}}
\newcommand{\p}{\mathbf{p}}

\newcommand{\PD}{\mathcal{P}}
\newcommand{\Pa}{\mathcal{P}^{\rm a}}
\newcommand{\Pb}{\mathcal{P}^{\rm b}}

\newcommand{\ClHMC}{$\mathcal{C}\mathrm{\ell}$HMC\xspace}

\newcommand{\python}{\textsc{Python}}
\newcommand{\mpipy}{\textsc{MPI4Py}}

\lstdefinestyle{DOS}
{
    backgroundcolor=\color{black},
    basicstyle=\scriptsize\color{white}\ttfamily
}

\newenvironment{rcases}
  {\left.\begin{aligned}}
  {\end{aligned}\right\rbrace}
  
\providecommand{\keywords}[1]{\noindent\textbf{\textit{Keywords: }} #1}    
  
\begin{document}

   \title{Medical Image Retrieval Based On the Parallelization of the Cluster Sampling Algorithm}

    \author[1]{Hesham Arafat Ali}
    \author[1]{Salah Attiya}
    \author[2]{Ibrahim El-henawy}
    \affil[1]{Computer Engineering and Systems Department\\ Faculty of Engineering\\ Mansoura University.}
    \affil[2]{Computer Science Department\\ Faculty of Computers and Informatics\\ Zagazig University}
   \date{}

\maketitle

\abstract{
  Cluster sampling algorithm is a scheme for sequential data assimilation developed to handle general non-Gaussian and nonlinear settings.
  The algorithm relaxes the Gaussian prior assumption widely used in the data assimilation context to approximate the prior distribution obtained by integrating 
  the posterior distribution in previous assimilation cycles. The algorithm can be in general used to solve inverse problems even when linearity or Gaussianity assumptions fail.
  The cluster sampling algorithm can be used to solve a wide spectrum of problems that requires data inversion such as image retrieval, tomography, weather prediction amongst others.
  In this paper, we develop parallel cluster sampling algorithms, and show that a multi-chain version is embarrassingly parallel, and can be used efficiently for medical image retrieval amongst other applications.
  Moreover, we present a detailed complexity analysis of the prposed parallel cluster samplings scheme and discuss their limitations.
  Numerical experiments are carried out using a synthetic one dimensional example, and a medical image retrieval problem.
  The experimental results show the accuracy of the cluster sampling algorithm to retrieve the original image from noisy measurements, and uncertain priors.
  Specifically, the proposed parallel algorithm increases the acceptance rate of the sampler from $44\%$ to $93\%$ with Gaussian proposal kernel, 
  and achieves an improvement of $29\%$ over the optimally-tuned Tikhonov-based solution for image retrieval.
 
  The parallel nature of the prposed algorithm makes the it a strong candidate for practical and large scale applications.
  }


\section{Introduction}\label{Sec:Introduction}

  Signal retrieval from noisy measurements (observations) involves solving an inverse problem.
  Inverse problems are essential in many fields such as image reconstruction or retrieval, tomography, weather prediction, and other predictions based on space-time models.
  The solution of inverse problems in case of space-time models usually employs a data assimilation (DA)\cite{Kalnay_2002_book,attia2015hmcsmoother,attia2016reducedhmcsmoother} methodology.
  DA refers to the process of fusing information about a physical system obtained from different sources in order to produces more accurate conclusions about the physical system of concern.

  Two approaches are widely employed to solve an inverse problem. 
  The first approach is a variational approach that involves solving an optimization problem with a regularized solution.
  The second approach is the statistical formulation of the DA problem which incorporates a prior distribution that encapsulates the knowledge about the system produced by the model, 
  prior to the incorporation of any other source of information.
  Given the prior information, a likelihood function, the posterior is formulated as best estimate of the truth.

  Markov Chain Monte-Carlo (MCMC) is one of the most powerful simulation techniques for sampling a high-dimensional probability distribution, given only it's shape function, without the intrinsic 
  need to the associated scaling factor. 
 
  HMC sampling filter~\cite{attia2015hmcfilter} is an accelerated Markov chain Monte-Carlo (MCMC) algorithm for solving the non-Gaussian  sequential DA ``filtering problem''. 
  This algorithm works by sampling the posterior distribution to produce description of the system state along with associated uncertainty. 
  Specifically, these algorithms follow a Hamiltonian Monte-Carlo (HMC) approach to sample the posterior.
 
  Cluster sampling filters (\ClHMC, and MC-\ClHMC)~\cite{attia2016clusterHMC} are developed as extension of the Hamiltonian Monte-Carlo (HMC) sampling filter presented in~\cite{attia2015hmcfilter} where
  the true (unknown) prior distribution is approximate using a Gaussian mixture model (GMM).

  Given the current computational power, it is natural to try to run Monte-Carlo simulations in parallel.
  However, Markov chains in general have to satisfy the so-called the ``{\it Markovian}" property which makes the chain generation an inherently sequential problem. 
  This restriction is mainly posed by the transition density function used to generate a proposal state given the current state of the chain.

  Two approaches have gained wide popularity to parallelize MCMC samplers. The parallel-chain approach proceeds by running several chains in parallel from different initial states. 
  The main disadvantage of this approach is that the burn-in stage has to be carried out by all chains independently, which limits the efficiency gained by running the chains on different processors. 
  Another difficulty with this approach is the aggregation of samples generated on different processors such that the combined ensemble correctly represents the mass of the target distribution.
  The second approach is to parallelize a single chain. 
 
  The parallel chain approach turns out to be surprisingly effective in practice. Moreover, if sufficient information about the geometry of the target distribution is available, we can guide the parallel 
  chains to sample effectively from the target distribution.

  The accuracy of \ClHMC filters to handle nonlinearity in both model dynamics, and observational mapping operator, puts it on the right direction of applicability to practical problems.
  The cost of serial \ClHMC is nearly similar to the cost of the original HMC sampling filter, however the MC-\ClHMC algorithm is naturally parallelizable. 
  
  Following a Bayesian approach, \ClHMC algorithm can be easily modified and applied for image retrieval given noisy image and a probabilistic representation of prior knowledge.
  This can be very useful in settings where several medical snapshots are collected for the same object, e.g. a tumor, of different resolution or uncertainty levels.
  
  Mathematical regularization is amongst the most popular methods for image reconstruction from noisy sources~\cite{ying2004tikhonov}.
  Among the regularization methods, the Tikhonov scheme is most popular due to the Gaussianity assumption about data noise, and the easiness to incorporate prior information.
  Disbite simplicity, the perfomace of this approach is highly influenced by the choice of the regularization parameter.
  
  Widely used methodologies for solving the Bayesian image retrieval problem include the algorithms discussed in~\cite{stathopoulos2011bayesian,hsiao2003bayesian}.
  In~\cite{hsiao2003bayesian}, the authors investigate statistical image reconstruction (SIR) with regularization based on the Markov random field (MRF) model.
  
  While, regularization approach is popular, it is sensitive to the choice of the regularization parameter. 
  
  Our main interest here is to develop highly accurate parallel Bayesian sampling algorithms that can be efficiently used for solving large-scale inverse problems, 
  and show that they are suitable for a wide spectrum of applications including medical image retrieval.
  
  In this work, we develop parallel cluster sampling algorithms, and show that a multi-chain version is embarrassingly parallel, 
  and can be used efficiently for medical image retrieval amongst other applications.
  The approach discussed in this work does not require regularization, and is designed to work in both Gaussian, and non-Gaussian case, where the computationl expense is minimized via parallelization.
  Specifically, in this paper, we focus on describing the complexity analysis of a specific scenario where the MC-\ClHMC is parallelized by running several chains in parallel to sample the posterior distribution.
  The algorithm proceeds by running several Markov chains in parallel such that the number of chains is specified by the the number of components in the mixture model.
  We will focus on the case where an ensemble of states, generated from an unknown prior distribution is available, and the likelihood function relating observations to target states is either a linear or a nonlinear map. 
  The prior distribution is approximated using a Gaussian mixture distribution which parameters are approximated based on the given prior ensemble by running an expectation maximization (EM) algorithm.

  In Section~\ref{Sec:Regularized_Image} we review the general iterative and Bayesian frameworks for inverse problems and image reconstruction.
  Section~\ref{Sec:Cluster_HMC} formulates the problem, and reviews the \ClHMC filter formulation.
  In Section~\ref{Sec:Parallelizing_ClHMC} we discuss opportunities for parallelization of \ClHMC.
  Section~\ref{Sec:Complexity_Analysis} presents a detailed complexity analysis of the proposed parallel version of \ClHMC filter.
  Numerical results are presented in Section~\ref{Sec:Numerical_Results}.
  Conclusions and future works are drawn in Section~\ref{Sec:Conclusions}

\section{Iterative and Bayesian Image Reconstruction}\label{Sec:Regularized_Image}
  As mentioned in Section~\ref{Sec:Introduction}, one of the most popular iterative reconstruction algorithms is Regularization-based algorithms.
  For the sake of completeness, we review the Tikhonov regularization approach~\cite{ying2004tikhonov,peterlik2008regularized} next, then we present the Bayesian formulation.
  
  The Tikhonov regularization approach involves solving the following optimization problem:
  \begin{equation}
  \label{eqn:Tikhonov}
    \xa = \min_{\x}{T(\x)} = {\lVert \mathcal{H}(\x) - \y \rVert }^2 _{R^{-1}} + \alpha \lVert \x \rVert^2 _ \mathbf{C},
  \end{equation}
  where $\alpha$ is the regularization parameter, and $\mathbf{C}$ is the regularization matrix and it can be chosen in many clever ways.
  Here $\mathcal{H}$ is an observation operator that maps the model space to the observation space. If the target state is directly observed then $\mathcal{H}=\mathbf{I}$, 
  where $\mathbf{I}$ is the identity operator.
  \noindent The weighted norm in Equation~\eqref{eqn:Tikhonov} is described as follows:
  \begin{equation} \label{eqn:weighted_norm}
    \mathbf{ \lVert c - d  \rVert }^2 _{M}  = \mathbf{ (c-d)^T M (c-d)   }
  \end{equation}
  The traditional approach to regularization is the variational formulation in which equation \eqref{eqn:Tikhonov} is minimized w.r.t $\x$.
  Usually, derivative-based iterative minimizaiton algorithms are employed to solve the problem described by~\eqref{eqn:Tikhonov}.
  The derivative of the objective function $T(\mathcal{H}(\x))$ w.r.t the parameter $]x$ is given by:
  \begin{equation} \label{eqn:Tikhonov_derivative}
    \nabla_{\x}{ T(\mathcal{H}{\x}) } = \left[ \partial \mathcal{H}(\x) \right]^{*} { \mathbf{R}^{-1} } ( \mathcal{H}(\x) - \y )  + \alpha \mathbf{C} \x,
  \end{equation}
  where ${ \left[ \partial \mathcal{H} \right]} ^{*}$ is the adjoint of the derivative, e.g. the Jacobian-transpose, of the observation operator $ \mathcal{H}$.
  In the case of a linear observaiton operator this is simply the transpose of the observation operator.
  
  In the statistical approach we infer the underlying state $\x$ based on a formulation constructed using Bayesian theory, where the goal is to represent the 
  state as a random variable which distribution is of interest.
  Assume $\Pb(\x)$ is a prior probability density function (PDF) representing prior knowladge about the state $\x$. 
  Assume also that $\PD(\y | \x)$ is the data likelihood function that describes the observational error distribution. 
  Using Bayes' theorem, the probability of the state $\x$ given the collected meauserments is characterized by the posterior distribution:
  \begin{equation} \label{eqn:posterior}
      \PD(\x | \y) \propto \PD(\y | \x ) \PD(\x)\,.
  \end{equation}

  A common practice is to assume that the prior distribution is, generally speaking, a multivariate normal (Gaussian) distribution centered around a forecasted
  or a first-guess state $\xb$, and have a predefined covariance structure $\mathbf{C}$, i.e. $\x \sim \mathcal{N}(\xb, C)$.
  
  For Gaussian priors and consequently Gaussian posteriors, the variational approach corresponds to finding the maximum aposterior estimate (MAP) of the posterior PDF.
  The MAP is the maximizer of the posterior PDF, or equivalently, the minimizer of its negative logarithm $ - \log {\left( \PD(\x | \y) \right)} $.
  Following Tikhonov regularization approach \eqref{eqn:Tikhonov}, and assuming Gaussian noise, the likelihood function reads:
  \begin{equation} \label{eqn:likelihood}
    \PD(\y | \x) \propto \exp{ \left( -\frac{1}{2} \lVert \mathcal{H}(\x) - \y \rVert^2 _{\mathbf{R^{-1}}} \right)} \,.
  \end{equation}
  Without loss of generality, if we assume $\x \sim \mathbf{N}(0, \mathbf{C})$, the prior would be on the form
  \begin{equation} \label{eqn:prior}
    \PD(\x) \propto \exp{ \left( - \frac{1}{2} \lVert \x\rVert^2 _{\mathbf{B}} \right) }\,,
  \end{equation}
  where $\mathbf{B}$ is the precision matrix, i.e. the inverse of the covariance $\mathbf{B}=\mathbf{C}^{-1}$. The MAP estimator in this formulation is the minimizer of 
  \begin{equation} \label{eqn:posterior_log}
    - \log{ \PD( \x | \y ) } \propto  \frac{1}{2} \lVert \mathcal{H}(\x) - \y \rVert^2 _{\mathbf{R^{-1}}}  + \frac{1}{2} \lVert \x\rVert^2 _{\mathbf{B}} \,.
  \end{equation}
  This shows the equivalence between Tikhonov regularization approach with the Bayesian formulation in the Gaussian linear settings.
  
  In the Bayesian approach, once the posteiror is constructed, a sampling mechanism is usually employed to estimate all the desired statistics of the posterior PDF, 
  such as the posteiror mean $\mathbb{E}(\x|\y)$ that can be used as a reliable estimate of the state given the data. Moreover, the generated ensemble can be used
  to estimate the posteiror covariance that can be used as a proxy prior error covariance for future applications of the inverse problem.
  Sampling the posterior PDF is usually carried out following a Monte-Carlo approach. 
  The most powerful Monte-Carlo sampling methodology is the general family Markov-Chain Monte Carlo (MCMC) samplers. 
  Sampling high dimensional distribution however is a very expensive process, and requires parallel efficient implementation to be considered practical.
  As explained in the next Section, MCMC is not limited to Gaussian or linear settings, and can be very efficient if implemented in parallel.

\section{Cluster Sampling Filter}\label{Sec:Cluster_HMC}
     
      Let $\x \in \mathbb{R}^{\nvar}$ is a discretized approximation of the true state of the model, for example the entensities of an image pixels.

      The prior distribution $\Pb(\x)$ encapsulates the knowledge about the system state before additional information is incorporated.
      The likelihood function $\PD(\mathbf{\y}|\x)$ quantifies the deviation of the prediction of model observations from the collected measurements $\y \in \mathbb{R}^{\nobs}$, 
      where $\nobs \leq \nvar$.

	  From Bayes' theorem, the posterior distribution $\Pa(\x) $ reads:
      \begin{equation} 
      \label{eqn:Bayes_Filtering_Rule}
       	\Pa(\x) = \PD(\x|\y)    = \frac{\PD(\y|\x) \Pb(\x) }{\PD(\y) } \propto \PD(\y|\x) \Pb(\x) \,,
      \end{equation}
      where $\Pb(\x)$ is the prior distribution, $\PD(\y|\x)$ is the likelihood function. $\PD(\y)$ acts as a scaling factor and is ignored in in the MCMC context. 

      Assuming the prior distribution is approximated by a Gaussian Mixture distribution, the prior takes the form:
      \begin{equation}
      \label{eqn:GMM_Filter_Prior}
	  \Pb(\x) = \sum_{i=1}^{\nc}{\tau_i\, \mathcal{N}(\mu_i,\, \Upsigma_i) }\,  
		  = \sum_{i=1}^{\nc}{\tau_i\,  \frac{(2\pi)^{-\frac{\nvar}{2}}}{\sqrt{|\Upsigma_i|}}\, 
				\exp{\left( -\frac{1}{2}  \lVert \x-\mu_i \rVert^2_{\mathbf{\Upsigma_i}^{-1}} \right)} }\,,
      \end{equation}
      where the weight $\tau_i$ quantifies the probability that an ensemble member $\x[e]$ belongs to the $i^{th}$ component, and $(\mu_i,\, \Upsigma_i)$ are the mean and the covariance matrix associated with the $i^{th}$ 
      component of the mixture model. Here $\x \in \mathtt{R}^{\nvar}$, where $\nvar$ the dimension of the target state space.
	      
      Assuming the observation errors are characterized by a Gaussian distribution $\mathcal{N}(0,\mathbf{R})$, the likelihood reads:
      \begin{equation}
      \label{eqn:Filtering_Likelihood}
	    \PD(\y|\x) = \frac{(2\pi)^{-\frac{\Nobs}{2}}}{\sqrt{|\mathbf{R}|}}\, \exp{\left( -\frac{1}{2} \lVert 
						  \mathcal{H}(\x) - \y \rVert^2 _{ \mathbf{R}^{-1}}\right)}\,,
      \end{equation}
      where $\mathbf{R}\, \in \mathtt{R}^{\Nobs\times\Nobs}$ is the observation error covariance matrix, and $\mathcal{H}:\mathtt{R}^{\nvar} \rightarrow \mathtt{R}^{\Nobs}$ is the observation operator that maps the 
      state space to the observation space. Here $\y \in \mathtt{R}^{\Nobs}$ is the observation vector. 
	  
      From Equations \eqref{eqn:Bayes_Filtering_Rule}, \eqref{eqn:GMM_Filter_Prior}, \eqref{eqn:Filtering_Likelihood},  the posterior takes the form:
      \begin{subequations}
      \label{eqn:Filtering_Posterior}
      \begin{align}
	    \Pa(\x) 
		    &=
			\frac{(2\pi)^{-\frac{\Nobs}{2}}}{\sqrt{|\mathbf{R}|}}\, \exp{\left( -\frac{1}{2}  \lVert \mathcal{H}(\x) -\y \rVert^2 _{\mathbf{R}^{-1} } \right) } \\ 
		    &           \hspace{2cm}\sum_{i=1}^{\nc}{\tau_i\,  \frac{(2\pi)^{-\frac{\nvar}{2}}}{\sqrt{|\Upsigma_i|}}\, \exp{\left( - \frac{1}{2}  \lVert \x-\mu_i \rVert^2_ {\mathbf{\Upsigma_i}^{-1}}  \right) } } \nonumber \\
		    &\propto 
		      \exp{\left( -\frac{1}{2}  \lVert \mathcal{H}(\x) -\y \rVert^2 _{\mathbf{R}^{-1} } \right) }
				\sum_{i=1}^{\nc}{\frac{\tau_i}{\sqrt{|\Upsigma_i|}}\, 
								  \exp{\left( -\frac{1}{2} \lVert \x-\mu_i \rVert^2_{\mathbf{\Upsigma_i}^{-1}} \right) }}  
      \end{align}
      \end{subequations}
      This mixture distribution may not correspond to a Gaussian mixture in general if the observation operator is a nonlinear map. 
      
      The negative-logarithm (negative-log) of the posterior distribution kernel~\eqref{eqn:Filtering_Posterior} is required by the HMC sampling algorithm.
      Specifically, the posterior negative-log is viewed as the potential energy function in the extended Hamiltonian phase space.
      The posterior negative-log is given by:
      \begin{equation}
      \label{eqn:Potential_Energy}
	\mathcal{J}(\x) = \frac{1}{2}  \lVert \mathcal{H}(\x) -\y \rVert^2 _{\mathbf{R}^{-1} } 
			  - \log{ \left(
					  \sum_{i=1}^{\nc}{\frac{\tau_i}{\sqrt{|\Upsigma_i|}}\, 
						  \exp{\left( -\frac{1}{2} \lVert \x-\mu_i \rVert^2_{\mathbf{\Upsigma_i}^{-1}} \right) }}
				  \right)}
      \end{equation}
      The derivative of the posterior negative-log reads:
      \begin{equation}
      \label{eqn:Potential_Energy_Gradient}
      \begin{aligned}
	\nabla_{\x} \mathcal{J}(\x) 
				    %
				    %
				    &= \mathbf{H}^T \mathbf{R}^{-1} ( \mathcal{H}(\x) -\y ) 
				      + \frac{ \sum_{i=1}^{\nc}{\frac{\tau_i}{\sqrt{|\Upsigma_i|}}\, 
								  \left( \exp{\left( -\frac{1}{2} \lVert \x-\mu_i \rVert^2_{\mathbf{\Upsigma_i}^{-1}} \right)  } \right) 
									  \mathbf{\Upsigma_i}^{-1} (\x-\mu_i)
								}
					      }{
						\sum_{i=1}^{\nc}{\frac{\tau_i}{\sqrt{|\Upsigma_i|}}\, 
							\exp{\left( -\frac{1}{2} \lVert \x-\mu_i \rVert^2_{\mathbf{\Upsigma_i}^{-1}} \right) }}
						}   \\
      \end{aligned}
      \end{equation}
      In this work, we sample the posterior distribution~\eqref{eqn:Filtering_Posterior} following a parallel chains approach given only ensemble of states generated from the prior distribution.

  \section{Parallelization of \ClHMC Filter}\label{Sec:Parallelizing_ClHMC}
    %
    In this section, we present a brief review of the MCMC sampling algorithm, and discuss opportunities of running \ClHMC filter on parallel architecture.
    We discuss the parallelism of \ClHMC filter even if the Hamiltonian system is replaced with a Gaussian proposal kernel build around the forecast.
    
    \subsection{Markov Chain Monte-Carlo (MCMC)}
      %
      MCMC is a sampling scheme capable of producing ensembles from an arbitrary distribution given it's shape function, without the need for the associated scaling factor. 

      The choice of the proposal kernel has the greatest influence on the performance of the sampler. 
      Here, we chose two proposal kernels; a) a Gaussian density function centered around the current state of the chain,  b) Hamiltonian Monte Carlo (HMC).
      The standard MCMC sampler is described in Algorithm~\ref{alg:mcmc}.
      \begin{center}
      \scalebox{0.99}{
      \begin{minipage}{1.0\linewidth}
	  \begin{algorithm}[H]
	  \SetKwInOut{Input}{Input}
	  \SetKwInOut{Output}{Output}
	  \label{alg:mcmc}
	  
	  \Input{An initial state for the chain ($\x^0$), and the proposal kernel $q$}
	  \Output{An ensemble of states from the posterior distribution $\propto \pi(\x)$}
	  
	  Initialize the chain to the state $ x^0 $  \;
	  Initialize k = 0 \;
	  \While {No sufficient samples are collected} {
	    Given the current state $\x^k$, use $q$ propose a state $x'$    \;
	    Calculate the acceptance probability (MH): $ a^k = \min {\left(1 \,,\,  \frac{ \pi(x') q(x', x^k) }{ \pi(x^k) q(x^k, x')} \right)}$  \;
	    Sample a uniform random number $u^k \sim \mathcal{U}(0,1)$ \;
	    \eIf { $ a^k > u^k$ } {
		    Accept the proposal: $ x^{k+1} = x' $ \;
	    }{
		    Reject the proposal: $ x^{k+1} = x^k $ \; 
	    }	      
	  }
	  \caption{Traditional MCMC algorithm to sample from $\pi(\x)$.}		
	  \end{algorithm}
	\end{minipage}%
	}
      \end{center}
      %
      %

      The standard MCMC algorithm\ref{alg:mcmc} generally suffers from random walk behaviour, slow convergence to the target density, low acceptance rate, and slow space exploration. 
      Moreover, the generated samples are highly correlated when the vanilla MCMC algorithm is used. 
      Many of these problems can be addressed by using Hybrid Monte Carlo (HMC). HMC uses a Hamiltonian system which plays the role of the proposal density.

      \subsection{The multi-chain MCMC algorithm (MC-MCMC)}
	%
	Although the traditional MCMC algorithm is inherently serial, there are several modifications that can be made to allow for parallelization. 
	In our approach, instead of constructing a single long Markov chain to produces $ \Nens $ samples, we generate several shorter Markov chains and divide the ensemble size $\Nens$ over these chains. 

	The constructed chains can run in parallel to sample different regions of the target distribution independently. 

	The parallel (MC-MCMC) sampler starts by running an Expectation Maximization step to build a Gaussian Mixture Model (GMM) approximation of the prior distribution.

	Once the GMM is constructed on the root processor, the GMM information is broadcasted to all the working nodes. Of course, if the number of processors is exactly the same as the number of components, each node is assigned one chain. 
	If the number of processors is less than the number of components/chains, we can assign several chains to each processor, e.g. based on the local ensemble sizes or simply in a round-robin fashion. 
	Caution has to be exercised to maintain a balanced load.	
	Once all chains have generated their assigned local samples, the ensembles are gathered to the root processor and returned as output. Of course parallel output can be considered to reduce the communication overhead.
	The steps of the proposed MC-MCMC scheme is detailed in Algorithm~\ref{alg:mc3}.
	\begin{center}
	  \scalebox{0.99}{
	  \begin{minipage}{1.0\linewidth}
	    \begin{algorithm}[H]
	    \label{alg:mc3}
		\SetKwInOut{Input}{Input}
		\SetKwInOut{Output}{Output}
		\Input{An ensemble of states from the prior distribution.}
		\Output{An ensemble of states from the posterior Mixture distribution~\eqref{eqn:Filtering_Posterior}.}	    
		    Run an EM algorithm (possibly in parallel) to build a GMM approximation of the prior distribution the given ensemble \;
		    \eIf{EM is run in parallel}{
			    \textsf{GatherAll} GMM information to all processors\;
		    }{
			    \textsf{Broadcast} GMM information to all processors.
		    }
		    $\nc$ chains are assigned to the available processors\;
		    The local ensemble size (sample size per chain) can be specified for example based on the weight of the prior weight of the corresponding component, multiplied by the likelihood of the mean of that component\;
		    Every chain is initialized to the mean of the corresponding component in the prior distribution\;
		    The parameters of the proposal kernel, e.g. covariance of the Gaussian kernel, or the mass matrix associated with HMC sampler, 
		    are set locally based on the statistics obtained from the prior ensemble under the corresponding component in the prior mixture.\;
		    After each chain collects it's assigned sample size,  \textsf{Gather} the ensembles generated by all nodes, and possibly weight them according to the importance of each component\;
		\caption{The MC-MCMC parallel sampling algorithm.}
		\end{algorithm}
	      \end{minipage}%
	      }
	    \end{center}

	    By running a Markov chain starting at each component of the mixture distribution, we ensure that the proposed algorithm navigates all modes of the posterior distribution, and covers all regions of high probability. 
	  
	    While parallelization of HMC itself can be considered for further increase in the performance, it has been avoided for clarity and to simplify the idea. 
	    We elected to use the Master-Slave parallel pattern, where a master core sends the information required for creating the individual chains. 

  \section{Complexity Analysis} \label{Sec:Complexity_Analysis}
    %
    In this section we provide a detailed theoretical discussion of the computational cost of the proposed parallel algorithm.
    Since sampling from a Gaussian distribution is essential for the two flavors tested here, we start with discussion the computational cost of sampling from a Gaussian distribution
    
    \subsection{Cost of sampling a Gaussian $\mathcal{N}(\mu,\, \Sigma)$}
     
      A scalar normal distribution can be sampled using many accurate algorithms such as the Mersenne Twister, Box-Muller transform, Marsaglia polar method, and Ziggurat algorithm.
      The least expensive is Ziggurat algorithm, where a typical value produced only requires the generation of one random floating-point value and one random table index, 
      followed by one table lookup, one multiply operation and one comparison.      
      The cost of generating an $\nvar-$dimensional standard normal random vector $\x\in\mathbb{R}^{\nvar}$ is $O(\nvar)$.
      To generate a multivariate normal (MVN) random vector $\emph{Ziggurat algorithm}$ from a general Gaussian distribution $ \mathcal{N}(\mu,\, \Sigma)$ the following steps are required:
      \begin{enumerate} 
	      \item Factorization of the covariance matrix $\Sigma$, to generate $\Sigma ^{\frac{1}{2}}$, e.g. using Cholesky decomposition
	      \item Draw a standard normal random vector $\y \in \mathcal{N}\mathbf{(0\, I}) $,
	      \item Scaling: $\x = \Sigma ^{\frac{1}{2}} \y + \mu$			 	
      \end{enumerate}
      If the covariance matrix $\Sigma$ is diagonal, the factorization costs $O(\nvar)$, while the cost of Cholesky decomposition in general is $O(\nvar^3)$. 
      If the covariance matrix $\Sigma$ is diagonal, the cost of the scaling step is $O(\nvar)$, otherwise the scaling cost is $O(\nvar^2)$.

      \subsection{Cost of MCMC with a Gaussian proposal density}
	
	  Assuming independent observations, evaluating the posterior PDF at a given state requires the evaluation of $\nc+1$ matrix-vector products. 
	  The cost of evaluating the posterior PDF is $O\left((\nc+1)\, \nvar\right) = O(\nvar)$ if the covariance matrices of the components the GMM prior are diagonal, otherwise, the cost is $O(\nvar^2)$.
	  
	  \paragraph{Cost of one MCMC step:}
	  %
	  In the presence of a Gaussian kernel, each step of the chain construction requires the following:
	  \begin{enumerate} 
		  \item One MVN random vector drawn from the Gaussian kernel,
		  \item Two function (posterior PDF) evaluations,
		  \item One draw from a uniform random distribution $\mathcal{U}(0,1)$,
		  \item One scalar comparison in the Metropolis-Hastings step.
	  \end{enumerate}

	  \noindent The cost of one MCMC step can be summarized as follows:
	  \begin{equation*} 
	    \resizebox{1.0\linewidth}{!}{ $
		  \begin{cases} 
			O(\nvar^2) + O(\nvar^2) + O(1)  =  O(\nvar^2)   &  \text{\small full GMM covariances, and full Gaussian kernel}  \\
			O(\nvar^2) + O(\nvar)   + O(1)  =  O(\nvar^2)   &  \text{\small full GMM covariances, and diagonal Gaussian kernel}      \\
			O(\nvar)   + O(\nvar^2) + O(1)  =  O(\nvar^2)   &  \text{\small diagonal GMM covariances, and full Gaussian kernel} \\
			O(\nvar)   + O(\nvar)   + O(1)  =  O(\nvar)     &  \text{\small diagonal GMM covariances, and diagonal Gaussian kernel} \\
			%
		      %
	      \end{cases}
	      $ }
	  \end{equation*}
	  
	  \noindent To find the total cost of MCMC sampling algorithm, we need to evaluate the total number of steps in the Markov chain. 
	  If the ensemble size is $\Nens$, we need to construct a chain of length $b_s + m_s \times \Nens$, where $b_s$ are burn-in steps carried out to achieve convergence to the stationary distribution,
	  and $m_s$ are mixing steps introduced to improve the sampler mixing and reduce the correlation between selected ensembles

	\paragraph{Total cost of MCMC sampling:}
	  The total cost of a serial MCMC with with a Gaussian proposal density reads:
	  \begin{equation*}
	    \resizebox{0.95\linewidth}{!}{ $
	    T_s = 
		      \begin{cases} 
			    O \Bigl( \left( b_s + m_s \, \Nens \right) \nvar   \Bigr)   &  \text{\small  diagonal prior covariances, and diagonal Gaussian kernel} \\
			    O \Bigl( \left( b_s + m_s \, \Nens \right) \nvar^2 \Bigr)   &  \text{\small  otherwise}
		  \end{cases}
		$ }
	  \end{equation*}
	  %
	  %

      \subsection{Cost of HMC}
	%
	The HMC sampler requires the evaluation of the gradient of the negative-log of the target distribution.
	\paragraph{Cost of one HMC step:}
	  %
	  In the case of HMC, each step of the chain construction requires the following:
	  \begin{enumerate} 
		\item One draw of a momentum $\p\sim \mathcal{N}(0,\mathbf{M})$ costs $O(\nvar)$ with diagonal $M$.
		\item Forward propagation of the pair $(\p,\, \x)$. 
	  \end{enumerate}
	  The cost of the second step is dominated by the cost of evaluating a Jacobian-vector product $O(\nvar^2)$. With step parameters $T=m\times h$, the cost is $O(m\,\nvar^2)$, where $m$ is the number of steps in the Hamiltonian trajectory.
	  Evaluating loss of energy requires evaluating the negative-log of the posterior shape function. Again, the cost of evaluating the posterior PDF is $O\left((\nc+1)\, \nvar\right) = O(\nvar)$ if the covariance matrices of the components 
	  the GMM prior are diagonal, otherwise, the cost is $O(\nvar^2)$.
	  
	  For Hamiltonian Monte Carlo (HMC), Assuming a diagonal mass matrix $\mathbf{M}$ the cost of one step of the chain is
	  \begin{equation}
	    \resizebox{0.95\linewidth}{!}{ $
		  \begin{cases} 
			O(\nvar) + O(m\, \nvar^2) + O(\nvar)   + O(1)  =  O(m\, \nvar^2)   &  \text{\small  diagonal prior covariances}     \\
			O(\nvar) + O(m\, \nvar^2) + O(\nvar^2) + O(1)  =  O(m\, \nvar^2)   &  \text{\small non-diagonal prior covariances} 
		  \end{cases}
		  $ }
	  \end{equation}
	  %
	      
	  \paragraph{Total cost of HMC sampling:}
	 
	    Following the discussion above, the cost of serial HMC sampler is $T_s = O \Bigl( \left( b_s + m_s \, \Nens \right)\,m\, \nvar^2   \Bigr)$.

	\subsection{Cost of MC-MCMC sampling}
	The serial complexity $T_s$ of MC-MCMC is summarized as follows:
	\begin{equation}
	    \resizebox{1.0\linewidth}{!}{ $
	T_s =           				
		\begin{cases}  
		   	\begin{rcases}
		      O \left( \left( b_s + m_s \, \Nens \right) \nvar   \right)\, ;\quad    &  \text{\begin{minipage}{15em} \small diagonal or spherical covariances of GMM, and proposal density \end{minipage} } \\
		      O \left( \left( b_s + m_s \, \Nens \right) \nvar^2 \right)\, ;\quad    &  \text{ \small  otherwise}\\ 
	        \end{rcases} \text{ \begin{minipage}{15em}\small Gaussian\\ proposal\end{minipage}} & \\
	          \hspace{1cm} \\
	          \begin{rcases}
		      	O \left( \left( b_s + m_s \, \Nens \right) \,m\,\nvar^2   \right)\, \quad\quad
 								      \end{rcases} \text{\begin{minipage}{10em}\small Hybrid Monte Carlo\end{minipage}} &
		\end{cases}
	  $ }
    \end{equation}
    %

	\paragraph{Parallel cost:}
      The parallel complexity can be studied clearly under a simplified (ideal) assumption, where each chain is to sample $\frac{\Nens}{\nc}$ ensemble points. 
      In this case, since we have $\nc$ chains the parallel cost (discarding the communication cost) of MC-MCMC is given by:
	\begin{equation}
	    \resizebox{1.0\linewidth}{!}{ $
		T_p =           				
				\begin{cases}  
					\begin{rcases}
				      O \left( \frac{\nc}{p} \left( b_s + m_s \, \frac{\Nens}{\nc} \right) \nvar   \right)\, ;\quad    &  \text{\begin{minipage}{15em} \small diagonal, or spherical covariances of GMM, and proposal density \end{minipage} } \\
				      O \left( \frac{\nc}{p} \left( b_s + m_s \, \frac{\Nens}{\nc} \right) \nvar^2 \right)\, ;\quad    &  \text{ \small  otherwise}\\ 
				\end{rcases} \text{ \begin{minipage}{15em}\small Gaussian\\ proposal\end{minipage}} & \\
				  \hspace{1cm} \\
				  \begin{rcases}
					O \left( \frac{\nc}{p} \left( b_s + m_s \, \frac{\Nens}{\nc} \right) \,m\,\nvar^2   \right)\, \quad\quad
				      \end{rcases} \text{ \begin{minipage}{15em}\small Hybrid\\ Monte Carlo\end{minipage}} &
			\end{cases}
		    $ }
	    \end{equation}
	%

	\paragraph{Speedup:}
	The speedup of MC-MCMC is given by:
	  \begin{equation}
	    \resizebox{0.6\linewidth}{!}{ $
		  S = \frac{T_s}{T_p} =  \frac{ \left( b_s + m_s \, \Nens \right)  }{\frac{\nc}{p} \left( b_s + m_s \, \frac{\Nens}{\nc} \right)  }      	
							    =  \frac{ p\,\left( b_s + m_s \, \Nens \right)  }{\nc \, \left( b_s + m_s \, \frac{\Nens}{\nc} \right)  }       				
	    $ }
	  \end{equation}
	  %

	\paragraph{Parallel efficiency:}
	The parallel efficiency of MC-MCMC is given by:
	  \begin{equation}
	    \resizebox{0.35\linewidth}{!}{ $
		  E = \frac{S}{p} =  \frac{ \left( b_s + m_s \, \Nens \right)  }{\nc \, \left( b_s + m_s \, \frac{\Nens}{\nc} \right)  }     				
	  $ }
	  \end{equation}
	  If we discard the burn-in stage, i.e. set $b_s=0$, the speedup, and the parallel efficiency simplify to:
	  \begin{equation*}
	    \resizebox{0.75\linewidth}{!}{ $
		  S = 
		  \begin{cases}
			      \p  \, &; \quad  p\leq \nc  \\
			      \nc \, &; \quad  p > \nc
			  \end{cases}
	      \,, \text{and,}\hspace{1.0cm}
		  E = \frac{S}{p} = 
		  \begin{cases}
			      1  \, &; \quad  p\leq \nc  \\
			      \frac{nc}{p} \, &; \quad  p > \nc 
			  \end{cases}
	      \,.
	      $ }
	  \end{equation*}
	  It follows that both speedup, and parallel efficiency are independent from the state space dimension $\nvar$.

	\paragraph{Communication overhead:}
	%
	Assuming serial GMM run on the root node, the cost of broadcasting GMM information to all nodes is the cost of broadcasting, the means, the covariance and/or precision matrices, and the weights.

	Assuming a linear communication model~\cite{kumar1994introduction}, and assuming that $t_s$, and $t_w$ are the startup, and the per-word transfer times respectively,
	the cost of broadcasting GMM information to $p$ nodes is given by:
	\begin{equation}
	  \resizebox{0.9\linewidth}{!}{ $
	    \begin{cases}
		      \Bigl( t_s + t_w \left(  ( 2 \nvar+1) \, \nc  \right) \Bigr) \log{(p)}
			\,;    &  \text{\begin{minipage}{12em} \small diagonal, or spherical GMM covariances \end{minipage} } \\
		      \Bigl( t_s + t_w \left(  ( \frac{\nvar^2}{\nc} + \nvar + 1) \, \nc  \right) \Bigr) \log{(p)}
			\,;    &  \text{\begin{minipage}{12em} \small tied GMM covariances \end{minipage} } \\
		    %
		      \Bigl( t_s + t_w \left(  ( \nvar^2 + \nvar + 1) \, \nc   \right) \Bigr) \log{(p)}	
			\,;    &  \text{\begin{minipage}{10em} \small full GMM covariances \end{minipage} }
	    \end{cases}
	  $ }
	\end{equation}
	%

	\noindent
	After sampling in parallel, the collected ensembles are gathered on the root node, at a cost
	$
	    \left( t_s + t_w \left(  \Nens \times \nvar   \right) \right) \log{(p)}\,.
	$
	%
      Consequently, the total communication cost reads:
      \begin{equation}
      \resizebox{0.9\linewidth}{!}{ $
	\begin{cases}
		  \Bigl( 2 t_s + t_w \left[  \left(  \left( \frac{\Nens}{\nc} + 2 \right) \nvar \, + 1 \right) \, \nc  \right] \Bigr) \log{(p)}      \,;    
																	&  \text{\begin{minipage}{12em} \small diagonal, or spherical GMM covariances \end{minipage} } \\
		  \Bigl( 2 t_s + t_w \left[  \left(  \left( \frac{\Nens+\nvar}{\nc} + 1\right) \nvar \, + 1 \right) \, \nc  \right] \Bigr) \log{(p)} \,;    
																	&  \text{\begin{minipage}{12em} \small tied GMM covariances \end{minipage} } \\
		%
		  \Bigl( 2 t_s + t_w \left[  ( \frac{\Nens \times \nvar}{\nc} + \nvar^2 + \nvar + 1) \, \nc   \right] \Bigr) \log{(p)}	
		    \,;    &  \text{\begin{minipage}{10em} \small full GMM covariances \end{minipage} }
	\end{cases}
	$ }
      \end{equation}
      %
      %

    \paragraph{Total parallel cost:}
    %
      It follows immediately from the discussion above, that 
     the total parallel cost of MC-MCMC sampling algorithm simplifies to:
	  \begin{equation} 
	  \resizebox{1.1\linewidth}{!}{ $
	    pT_p =       				
			  \begin{cases}  
			      &
			      \begin{rcases}
				  \begin{cases}  
				\Bigl( 2 t_s + t_w \left[  \left(  \left( \frac{\Nens}{\nc} + 2 \right) \nvar \, + 1 \right) \, \nc  \right] \Bigr)p \log{(p)}  + 
				O \left( \nc \left( b_s + m_s \, \frac{\Nens}{\nc} \right) \nvar   \right)\, ;\quad    &
														    \text{\begin{minipage}{15em} \small diagonal, or spherical covariances of GMM, and proposal density \end{minipage} } \\
				  \hspace{1cm} &\\
				\Bigl( 2 t_s + t_w \left[  \left(  \left( \frac{\Nens+\nvar}{\nc} + 1\right) \nvar \, + 1 \right) \, \nc  \right] \Bigr) p\log{(p)} +
				O \left( \nc \left( b_s + m_s \, \frac{\Nens}{\nc} \right) \nvar^2   \right)\, ;\quad    &
														    \text{\begin{minipage}{12em} \small tied covariances of GMM \end{minipage} } \\
				  \hspace{1cm} &\\
				\Bigl( 2 t_s + t_w \left[  ( \frac{\Nens \times \nvar}{\nc} + \nvar^2 + \nvar + 1) \, \nc   \right] \Bigr) p\log{(p)}	 +
				O \left( \nc \left( b_s + m_s \, \frac{\Nens}{\nc} \right) \nvar^2 \right)\, ;\quad   &   \text{ \small  otherwise}\\ 
			      \end{cases}
			  \end{rcases} \text{ \begin{minipage}{15em}\small Gaussian\\ proposal\end{minipage}} \\
			  & \vspace{1cm} \\
			  &
			  \begin{rcases}
				O \left( \nc \left( b_s + m_s \, \frac{\Nens}{\nc} \right) \,m\,\nvar^2   \right)\, + \,
				      \begin{cases}
						\Bigl( 2 t_s + t_w \left[  \left(  \left( \frac{\Nens}{\nc} + 2 \right) \nvar \, + 1 \right) \, \nc  \right] \Bigr) p\log{(p)}      \,;    
																				      &  \text{\begin{minipage}{13em} \small diagonal, or spherical GMM covariances \end{minipage} } \\
						\hspace{1cm} \\
						\Bigl( 2 t_s + t_w \left[  \left(  \left( \frac{\Nens+\nvar}{\nc} + 1\right) \nvar \, + 1 \right) \, \nc  \right] \Bigr) p\log{(p)} \,;    
																				      &  \text{\begin{minipage}{12em} \small tied GMM covariances \end{minipage} } \\
					\hspace{1cm} \\
						\Bigl( 2 t_s + t_w \left[  ( \frac{\Nens \times \nvar}{\nc} + \nvar^2 + \nvar + 1) \, \nc   \right] \Bigr) p\log{(p)}	
						  \,;    &  \text{\begin{minipage}{12em} \small full GMM covariances \end{minipage} }
				  \end{cases}
				  \quad
			      \end{rcases} \text{ \begin{minipage}{15em}\small HMC \end{minipage}} 
		      \end{cases}
	      $ }
	  \end{equation} 
	  %
	  %

	\paragraph{Total overhead:}
    %
          	%
The total overhead function ($T_o = P T_p -T_s$) of MC-MCMC reads:
	  \begin{equation} 
	  \resizebox{1.1\linewidth}{!}{ $
	    T_o = 
			  \begin{cases}  
			      &
			      \begin{rcases}
				  \begin{cases}  
				  \Bigl( 2 t_s + t_w \left[  \left(  \left( \frac{\Nens}{\nc} + 2 \right) \nvar \, + 1 \right) \, \nc  \right] \Bigr)p \log{(p)}  + 
				O \left( \left(\nc-1\right) b_s \, \nvar  \right)\, ;\quad    &
														    \text{\begin{minipage}{15em} \small diagonal, or spherical covariances of GMM, and proposal density \end{minipage} } \\
				  \hspace{1cm} \\
				  \Bigl( 2 t_s + t_w \left[  \left(  \left( \frac{\Nens+\nvar}{\nc} + 1\right) \nvar \, + 1 \right) \, \nc  \right] \Bigr) p\log{(p)} +
				O \left( \left(\nc-1\right) b_s \, \nvar^2   \right)\, ;\quad    &
														    \text{\begin{minipage}{12em} \small tied covariances of GMM \end{minipage} } \\
				  \hspace{1cm} \\
				  \Bigl( 2 t_s + t_w \left[  ( \frac{\Nens \times \nvar}{\nc} + \nvar^2 + \nvar + 1) \, \nc   \right] \Bigr) p\log{(p)}	 +
				O \left( \left(\nc-1\right) b_s \, \nvar^2 \right)\, ;\quad    &  \text{\small  otherwise}\\ 
			  \end{cases}
			  \end{rcases} \text{ \begin{minipage}{15em}\small Gaussian\\ proposal\end{minipage}}  \\
			  \vspace{1cm} \\
			  &
			  \begin{rcases}
				O \left( \left( \nc - 1  \right)\, b_s \,m\,\nvar^2   \right)\, + \,
							  \begin{cases}
								    \Bigl( 2 t_s + t_w \left[  \left(  \left( \frac{\Nens}{\nc} + 2 \right) \nvar \, + 1 \right) \, \nc  \right] \Bigr) p\log{(p)}      \,;    
																							  &  \text{\begin{minipage}{14em} \small diagonal, or spherical GMM covariances \end{minipage} } \\
								    \hspace{1cm} \\
								    \Bigl( 2 t_s + t_w \left[  \left(  \left( \frac{\Nens+\nvar}{\nc} + 1\right) \nvar \, + 1 \right) \, \nc  \right] \Bigr) p\log{(p)} \,;    
																							  &  \text{\begin{minipage}{14em} \small tied GMM covariances \end{minipage} } \\
							    \hspace{1cm} \\
								    \Bigl( 2 t_s + t_w \left[  ( \frac{\Nens \times \nvar}{\nc} + \nvar^2 + \nvar + 1) \, \nc   \right] \Bigr) p\log{(p)}	
								      \,;    &  \text{\begin{minipage}{12em} \small full GMM covariances \end{minipage} }
						      \end{cases}
				  \quad
			      \end{rcases} \text{ \begin{minipage}{15em}\small HMC \end{minipage}} 
		      \end{cases}
	      $ }
	  \end{equation} 
	  %
          	%

	\paragraph{Isoefficiency:}
	  %
	  Assuming the burn-in stage is discarded, i.e. set $b_s=0$.
	  With $\Nens\geq 2\nc$, the isoefficiency function $W(p) = \frac{E}{1-E}T_o = kT_o$ simplifies to (the dominant terms):
	  \begin{equation} 
	  \resizebox{0.95\linewidth}{!}{ $
	    W(W, p) = 
			  \begin{cases}  
				  k \, t_w\, \Nens \, \nvar \,   p \, \log{(p)}  \, ;\quad    &
														    \text{\begin{minipage}{18em} \small diagonal, or spherical GMM covariances, and {HMC, or Gaussian proposal with diagonal covariance} \end{minipage} } \\
				  \hspace{1cm} &\\
				  k \, t_w\, \left( \Nens+\nvar\right) \, \nvar \,   p \, \log{(p)}  \, ;\quad    &
														    \text{\begin{minipage}{16em} \small tied covariances of GMM,  and {HMC, or Gaussian proposal} \end{minipage} } \\
				  \hspace{1cm} & \\
				  k \, t_w\, \left( \Nens\,+\nvar\, \nc \right) \, \nvar \,   p \, \log{(p)}  \, ;\quad    &  \text{\begin{minipage}{16em} \small full covariances of GMM,  and {HMC, or Gaussian proposal} \end{minipage} }\\ 

		      \end{cases}
		$ }
	  \end{equation} 
	  %

\section{Numerical Experiments and Performance} \label{Sec:Numerical_Results}
 
  In this section we present numerical experiments to assess the complexity analysis provided in Section~\ref{Sec:Complexity_Analysis}.
  As mentioned above, the speedup, and parallel efficiency are independent of problem dimensionality.
  We discuss the computational cost and the performance using one dimensional examples.

  To execute the Markov chains in parallel, we used the \mpipy~package, which provides bindings of the Message Passing Interface (MPI) standard for the \python~programming language, 
  allowing any \python program to exploit multiple processors.

  we present numerical experiments to assess the complexity analysis provided in Section~\ref{Sec:Complexity_Analysis}.
  Specifically, we show numerical results for the parallel \ClHMC flavors discussed in this paper using one dimensional synthetic example, and a two dimensional image retrieval experiment.
  %
 
  \subsection{One-dimensional example:}
\begin{figure}[!htbp]
    \centering
    \subfigure[Gaussian Proposal, Serial MCMC]{%
	    \includegraphics[width=0.4\linewidth,height=3.0cm]{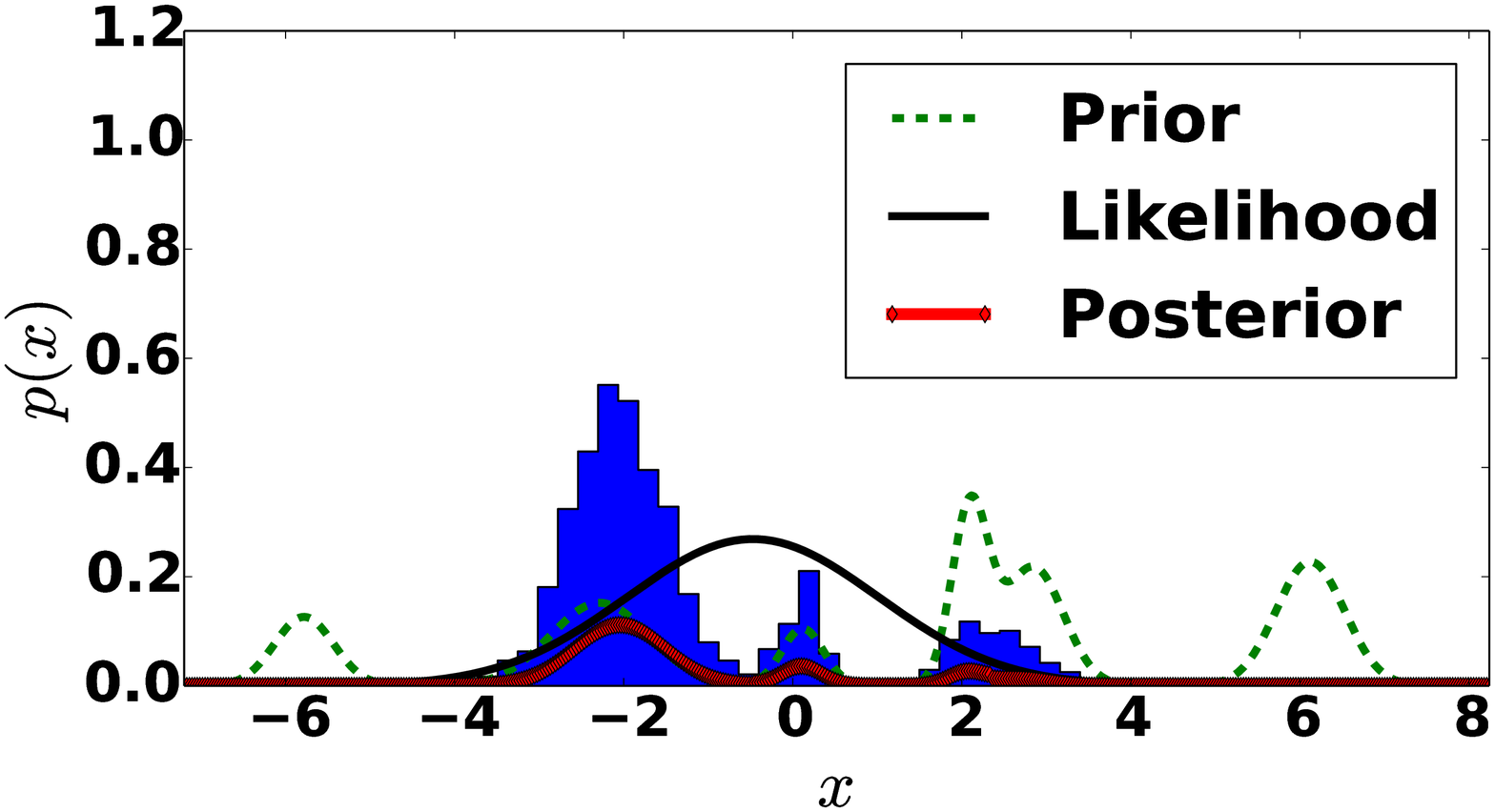}
	    \label{fig:Sample_oneD_linear_serial_Gaussian}
	    }
    \subfigure[Gaussian Proposal, Parallel MCMC]{%
	    \includegraphics[width=0.4\linewidth,height=3.0cm]{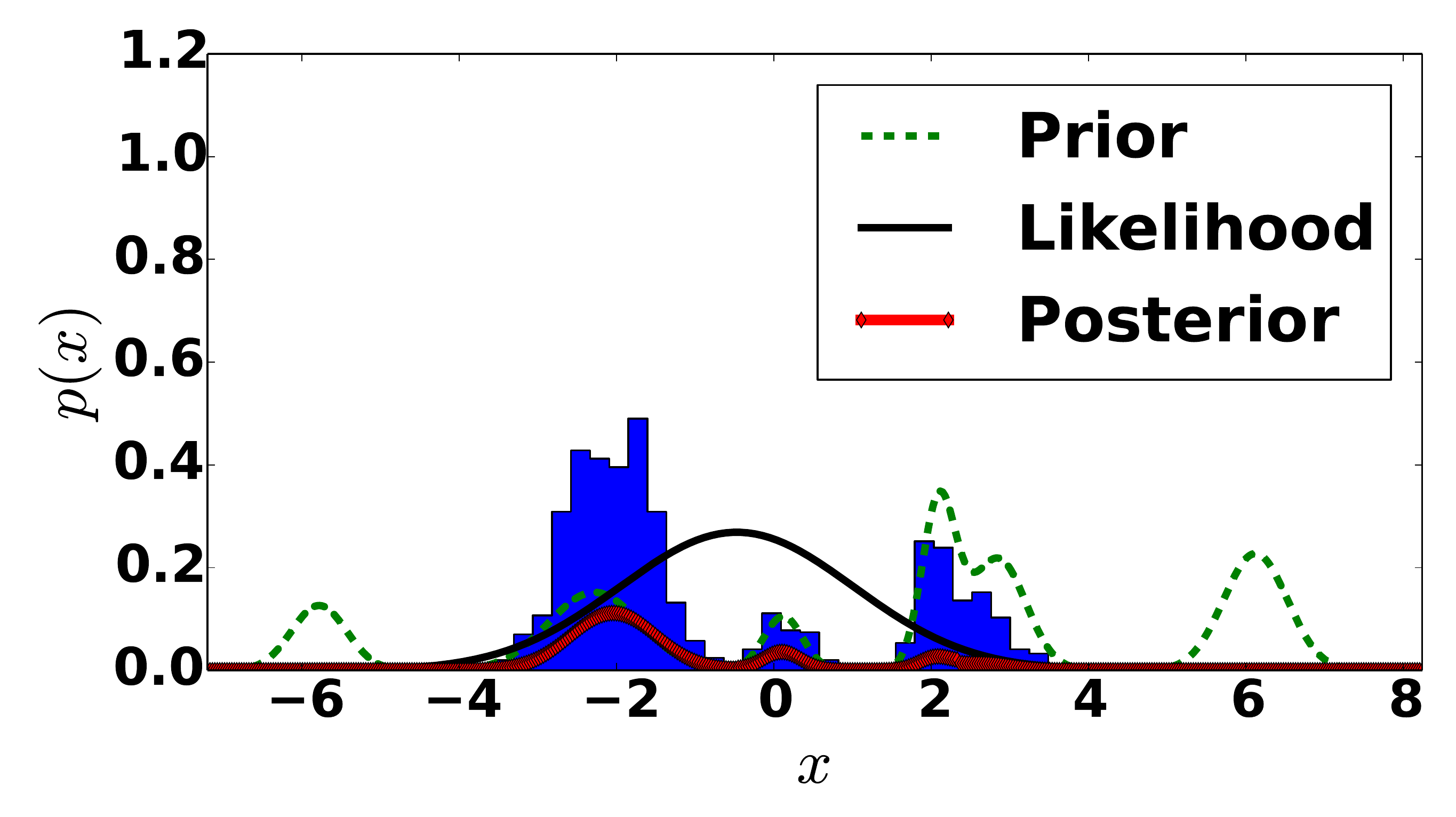}
	    \label{fig:Sample_oneD_linear_parallel_Gaussian}
	    }
    \hfill
    \subfigure[HMC Proposal, Serial MCMC]{%
	    \includegraphics[width=0.4\linewidth,height=3.0cm]{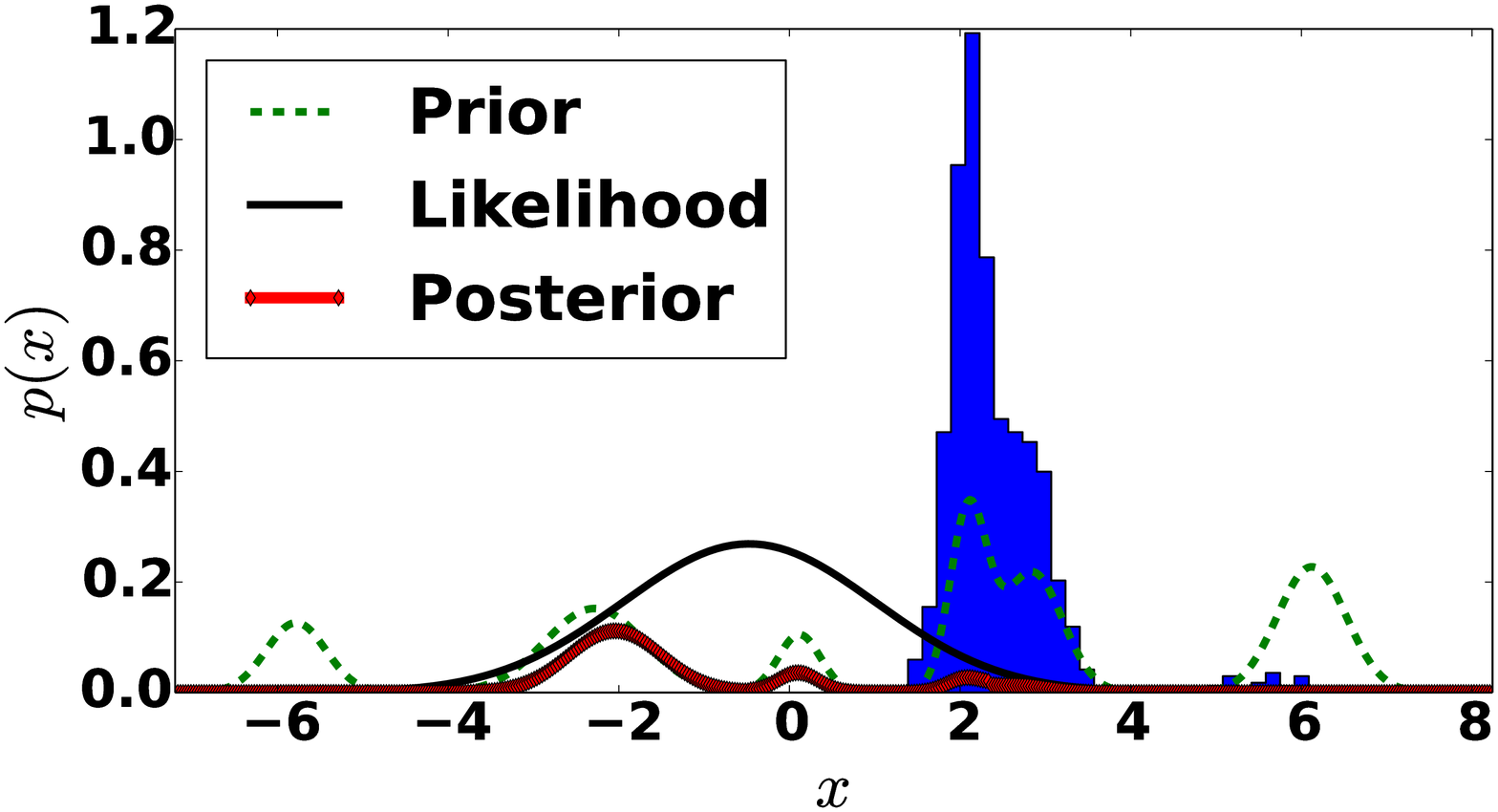}
	    \label{fig:Sample_oneD_linear_serial_HMC}
	    }
    \subfigure[HMC Proposal, Parallel MCMC]{%
	    \includegraphics[width=0.4\linewidth,height=3.0cm]{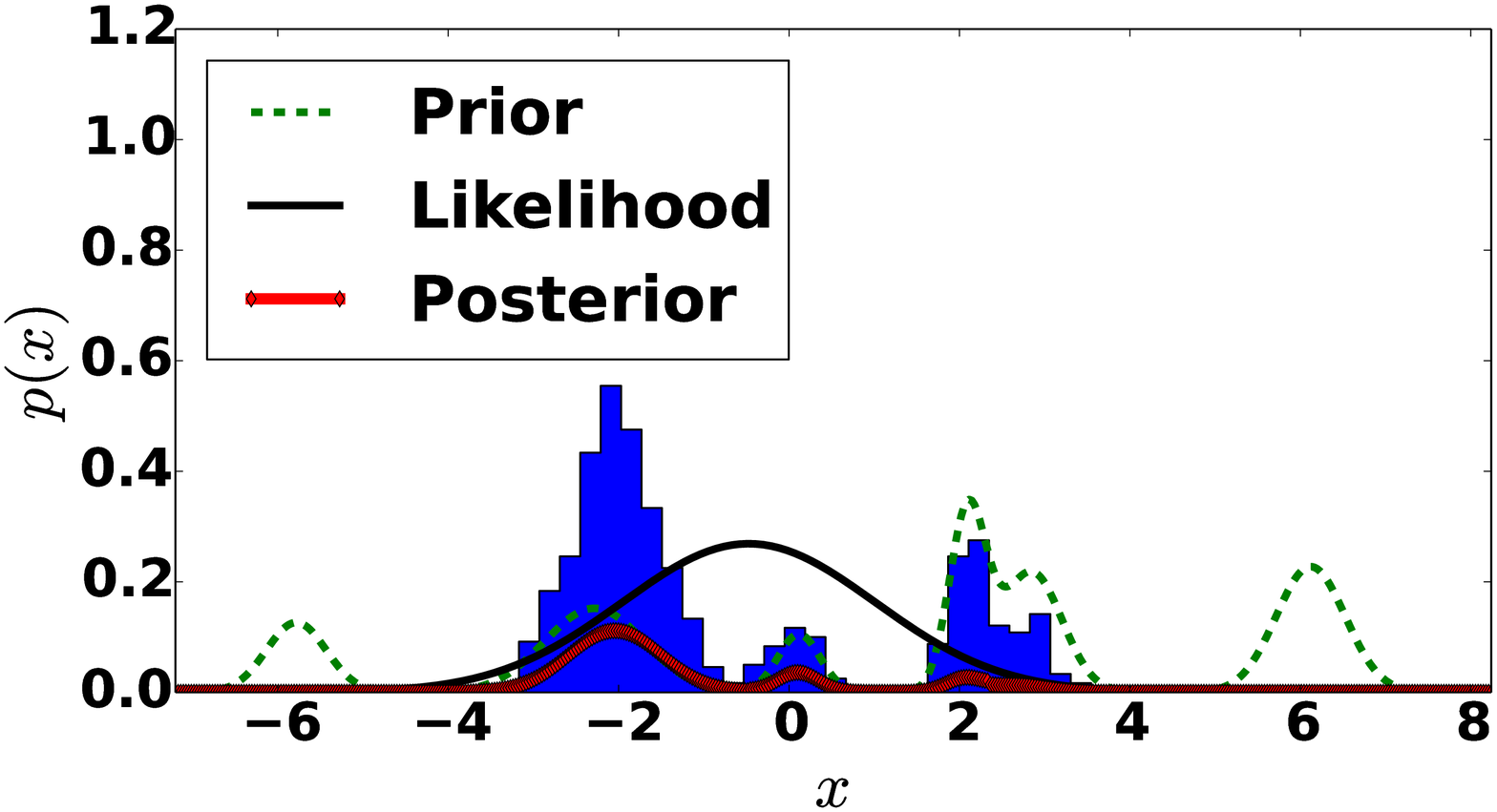}
	    \label{fig:Sample_oneD_linear_parallel_HMC}
	    }
    %
    %
    \caption{Sampling results of a one dimensional posterior with seven components. The results of the serial MCMC and MC-MCMC with two proposal mechanisms are shown. 
	      The mode (serial vs. parallel) and the proposals are shown under each panel. The ensemble size here is $1000$}
    \label{fig:Sampling_result1D}
    \end{figure}

    %
    Following the strategy described in~\cite{attia2016clusterHMC}, we start with a synthetic prior ensemble generated from a GMM with $\nc=5$.
    A GMM approximation of the true prior probability distribution is constructed using the EM algorithm .
    The model selection criterion used here is AIC.
    The parameters of the true GMM prior are: 
    \begin{equation} \label{eqn:OneD_Example_true_gmm_parameters}
      \begin{split}
	\{(\tau_i;\,\mu_i, \sigma_i^2)\}_{i=1,\dots,5}=\{(0.09;\, -6.0,  0.20),\, 
							  (0.19;\, -2.5,  0.28),\, 
							  (0.09;\,  0.0,  0.08),\,\\ 
							  (0.28;\,  2,5,  0.24),\,
							  (0.15;\,  6.0,  0.28),\, 
							  (0.15;\,  6.5,  0.08),\, \\
							  (0.03;\,  7.5,  0.12),\, 
							  (0.02;\,  8.0,  0.04) \, \} \,.
      \end{split}
    \end{equation}
    Assume the observation errors follow Gaussian distribution with zero mean, and variance $2.2$.
    Assuming a synthetic observation $\y=-1.0$, the observation likelihood function given by:
    \begin{equation}
    \centering
    \PD(\y|\x) =\frac{1}{\sqrt{2.2}\, \sqrt{ 2\pi }}\, 
    \,\exp{\Bigl( -\frac{1}{2}\, \frac{\left( \x -\y  \right)^2}{ 2.2 } \Bigr) }\,.
    \end{equation}
    The generated GMM approximation of the prior has  $\nc=7$ and the following parameters:
    \begin{equation} \label{eqn:OneD_Example_EM_gmm_parameters}
    \begin{split}
      \{(\tau_i;\,\mu_i, \sigma_i^2)\}_{i=1,\dots,7}=\{(0.111;\, -5.78, 0.123),\, (0.177;\, -2.49, 0.223),\, \\ (0.045;\, -1.49, 0.001),\,
	(0.065;\, 0.12, 0.061),\, (0.146;\, 2.05, 0.032),\, \\(0.225;\, 2.78, 0.148),\, (0.231;\, 6.12, 0.164) \, \} \,.
    \end{split}
    \end{equation}
    Figure~\ref{fig:Sampling_result1D} shows the results of sampling a one dimensional posterior~\eqref{eqn:Filtering_Posterior} with seven components. 
    Since we are mainly interested in asymptotic behaviour of the sampler, and to avoid the effect of sampling error, the ensemble size here is set to $1000$. 
    Despite the good mass distribution resulting from the use of a Gaussian density with the serial MCMC sampler~\ref{fig:Sample_oneD_linear_serial_Gaussian}, 
    the acceptance rate is $\approx 45 \%$ resulting in a large amount of wasted calculations.
    On the other hand, the parallel MC-MCMC with Gaussian proposal kernel~\ref{fig:Sample_oneD_linear_parallel_Gaussian} improves the acceptance rate to $\approx 81 \%$. 
    The acceptance rate is increased due to the local adjustment of the sampler hyperparameters based on the local ensemble under the corresponding prior component in the mixture.
    While the Gaussian-based MCMC sampler represents an acceptable mass distribution, it suffers from random walk behaviour leading to the demonstrated low acceptance rate.
    One the other hand HMC sampling results in general in high acceptance rate. 
    Unfortunately, as explained by the results in Figure~\ref{fig:Sample_oneD_linear_serial_HMC} is unable to sample all probability modes.
    The acceptance rate in the serial \ClHMC sampler here is $96\%$.
    By running \ClHMC sampling methodology in parallel, the acceptance rate drops to $94\%$ (which is still very high). However, the mass distribution is much better than the serial case.
    This is supported by results in Figure~\ref{fig:Sample_oneD_linear_parallel_HMC} compared to Figure~\ref{fig:Sample_oneD_linear_serial_HMC}.

\begin{figure}[!htbp]
    \centering
    \subfigure{%
	    \includegraphics[width=0.48\linewidth,height=3.0cm]{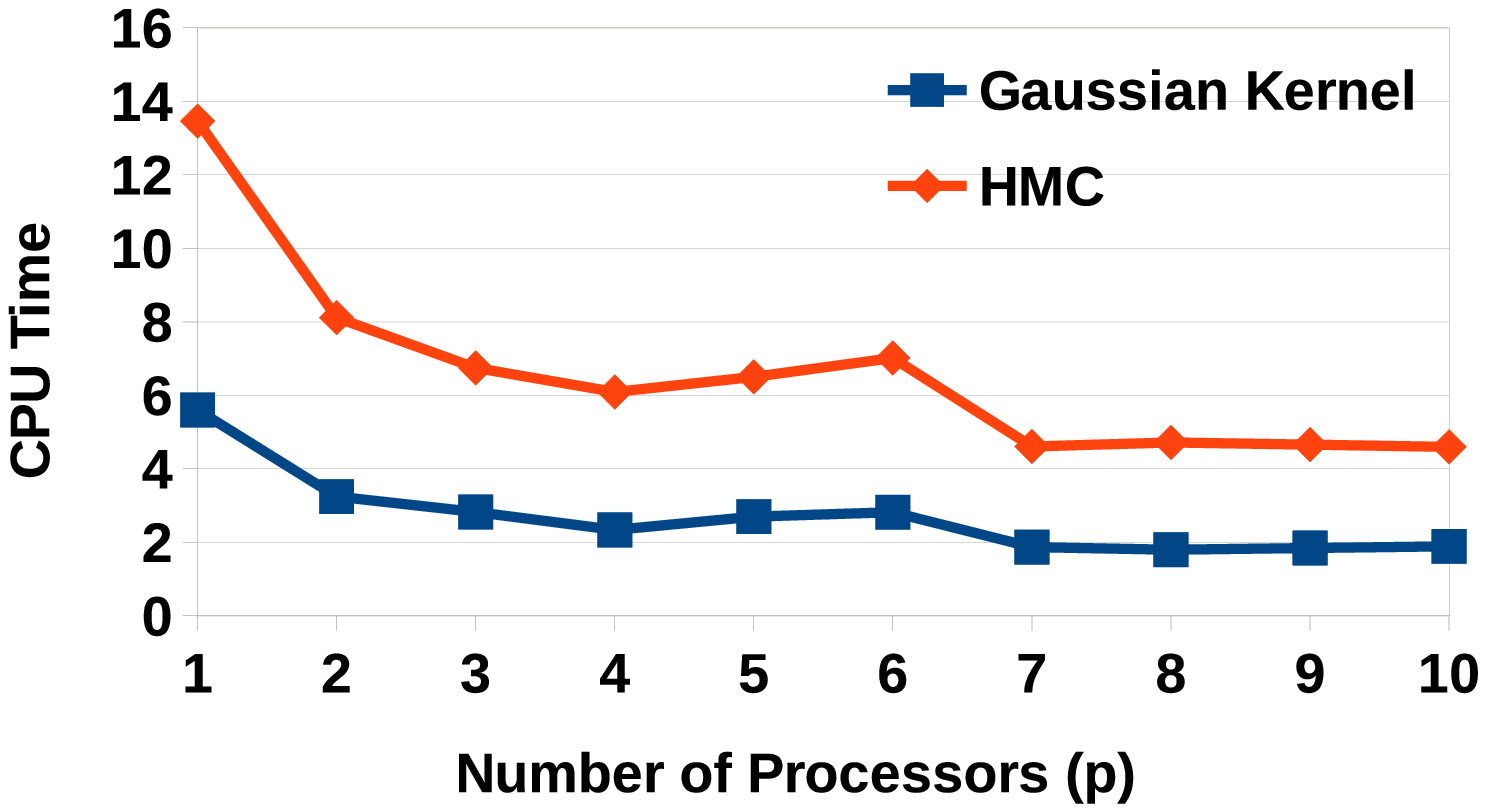}
	    }
    \hfill
    \subfigure{%
	    \includegraphics[width=0.48\linewidth,height=3.0cm]{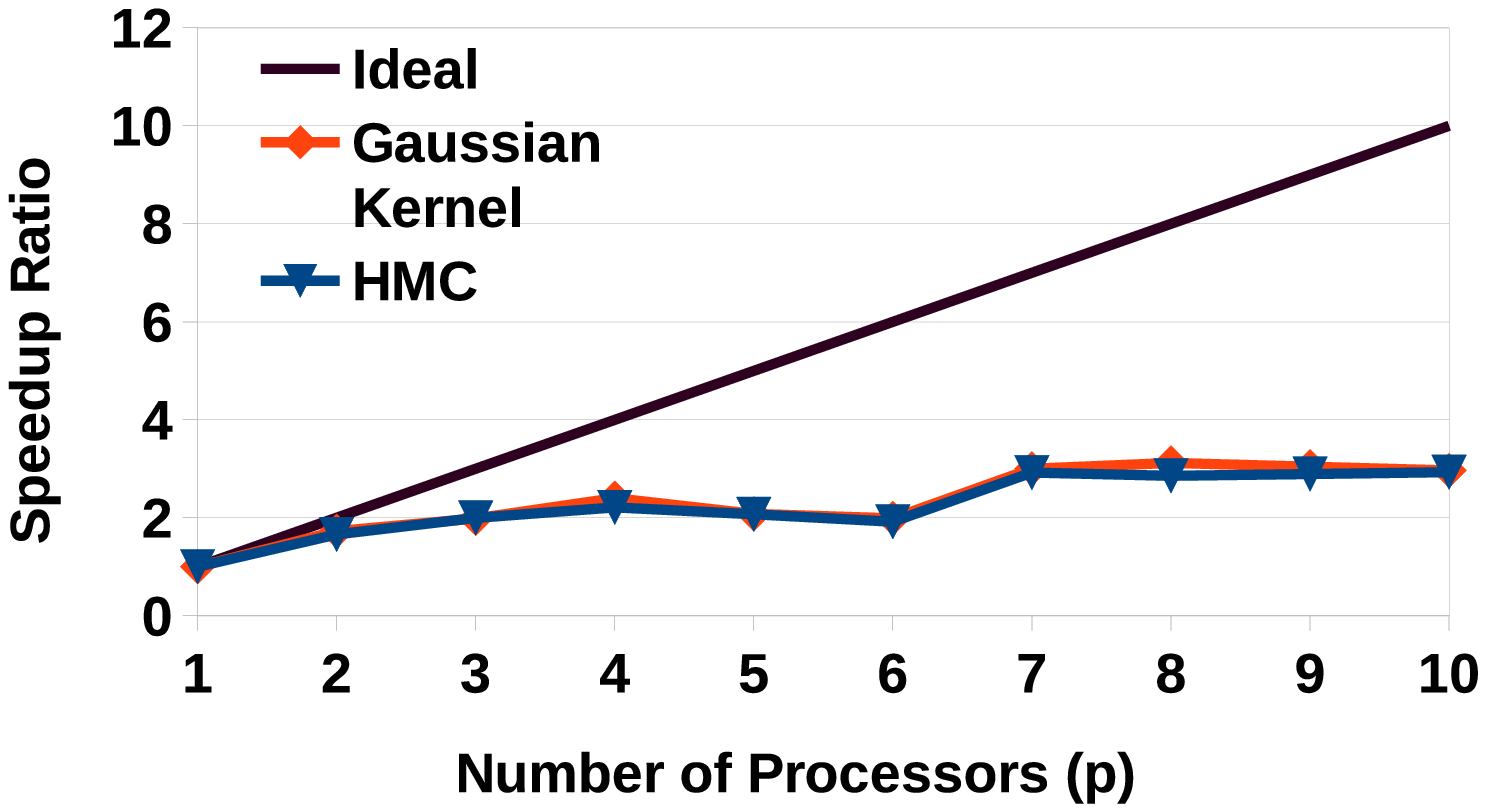}
	    }
    %
    %
    \caption{Sampling results of a one dimensional posterior with four components. The CPU time, and speedup results of MC-MCMC with two proposal mechanisms are plotted.
	      The ensemble size here is $1000$
	    }
    \label{fig:speedup}
    \end{figure}

    Figure~\ref{fig:speedup} shows the CPU time and speedup results of the clustering sampling algorithms with Gaussian and HMC proposal mechanisms.
    As suggested by the analysis in Section~\ref{Sec:Complexity_Analysis}, the CPU time becomes flat once the number of processors reaches $7$, the number of components in the mixture.
    %
    %
    %

\begin{figure}[!htbp]
    \centering
    \includegraphics[width=0.65\linewidth,height=3.0cm]{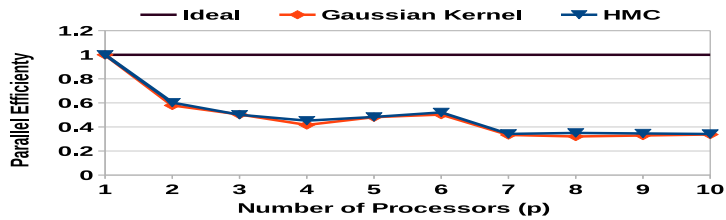}
    \caption{Sampling results of a one dimensional posterior with four components. The parallel efficiency results of MC-MCMC with two proposal mechanisms are plotted.
	      The ensemble size here is $1000$
	    }
    \label{fig:efficiency}
    \end{figure}

    The parallel efficiency results are shown in Figure~\ref{fig:efficiency}.
    The numerical results shown here suggest that, running the clustering filters in parallel not only results in computational saving, but also can potentially increase the sampling accuracy.
    While the discussed parallelization of the sampler reduces the computational time, the numerical results suggest that more parallelization effort is needed in order to achieve higher efficiency. 
    In the current settings, the chains are assigned to processes in a round-robin fashion.
    The performance of the sampler can be greatly enhanced if the a smart scheduler is used such that the parallel chains are assigned to processors based on the sample size per chain.
    %

    \subsection{Two-dimensional example}
    %
    Here we test the the paralle \ClHMC sampling algorithm as a tool for statistical medical image retrieval.
    We employ a non-linear Gaussian convolution filter as a forward operator $\mathcal{H}$. 
    The convolution filter is applied to a two-dimensional image resulting in a blurred image, then Gaussian noise is added to collect synthetic measured/observed image. 
    Here, the vector $\x$ represents intensities of the image pixels  arranged in a column vector.
    The observation nosie level is set to be $9\%$ of the average intensity of the original image.
    This formulation clearly results in a nonlinear inverse problem that can be challanging for traditional approaches such as Tikhonov regularization. 
    For , the Jacobian of the convoluted image with respect to the intensities of the image pixels is found to be a Toeplitz matrix.
    
    The goal here is to retrieve the original image given the noisy measurement, and a sample drawn from the probability distribution from which the blurred image is drawn.
    This is is a relevant problem description in many cases, where several low resolution or blurrd images are taken along with the collected measurement.
    
\begin{figure}[H]
    \centering
	\subfigure[Original Image.]{%
	      \includegraphics[width=0.270\linewidth]{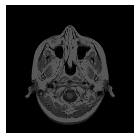}   
	    \label{fig:TwoD_Original}}
	\,
	\subfigure[Blurred Image.]{%
	      \includegraphics[width=0.275\linewidth]{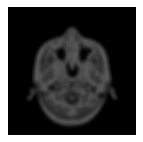}   
	    \label{fig:TwoD_Blurred}}
	\,
	\subfigure[Data: Blurred Image with added noise.]{%
	      \includegraphics[width=0.270\linewidth]{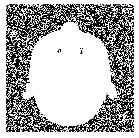}   
	    \label{fig:TwoD_Noisy}}
	    
	      \caption{Inputs to the sampling filter. Original image $\x$, blurred image $\mathcal{H}(\x)$, and noisy image $\y$.} 
	      \label{fig:TwoD_Inputs}
    \end{figure}

    Figure \ref{fig:TwoD_Inputs} shows the original (true) image, the blurred image constructed by the convolution filter, and the noisy image, i.e. the blurred image with additive noise.

    To create a synthetic non-Gaussian sample, we sampled $50$ images from a Gaussian distribution centered around the blurred image with variances equal to $8\%$ of 
    the average entensity of the original image.
    To create a synthetic prior sample, we have selected $\Nens=30$ uniformly random distributed images from the generated Gaussin sample. 
    This procedure is guaranteed to result in a non-Gaussian prior, and is powerful to test the GMM prior assumption.
    
    The \ClHMC sampling algorithm with both Gaussian and Hamiltonian kernels performed similarily. The acceptance rates however were quite different. Specifically, the rejection rate in the 
    case of Gaussian proposal was $56\%$ resulting in wasted computations, while HMC rejection rate was $7\%$.
    
    The initial state of each chain is chosen as the mean of the corresponding component in the prior mixture, and the parameters of the Hamiltonian system are tuned empirically to give 
    acceptable acceptance rate. $30$ sample members are collected from the posterior distribution.
    
\begin{figure}[H]
    \centering
      \subfigure[Tihonov regularized solution]{%
	    \includegraphics[width=0.293\linewidth]{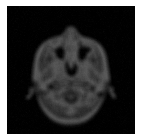}   
	    \label{fig:Tikhonov_regularized_sol}}
      \,
	\subfigure[\ClHMC: sample mean]{%
	    \includegraphics[width=0.2705\linewidth]{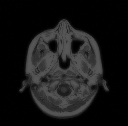}   
	    \label{fig:TwoD_sample_mean}}
      \,
      \subfigure[\ClHMC: sample mean]{%
	    \includegraphics[width=0.2705\linewidth]{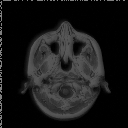}   
	    \label{fig:TwoD_sample_median}}
      
	    \caption{Inverse problem solution using Tihonov regularization and the parallel \ClHMC posterior sample. Mean and Median, of the collected $30$ sample members using paralle \ClHMC. The method and statistic used are shown under each panel.} 
	    \label{fig:TwoD_Retrieved}
    \end{figure}

    The mean and median of posterior samples collectecd using the parallel \ClHMC sampling algorithm are shown in Figure \ref{fig:TwoD_Retrieved} (panels \ref{fig:TwoD_sample_mean}, and \ref{fig:TwoD_sample_median}).
    For the sake of comparison to one of the most popular and widely used approaches, we show the results obtained using Tikhonov regularization approach with regularization parameter optimally tuned following an L-curve approach.
    The Tikhonov-regularized solution is shown in panel \ref{fig:Tikhonov_regularized_sol} of Figure \ref{fig:TwoD_Retrieved}.

    By comparing results in Figure~\ref{fig:TwoD_Retrieved}, to the blurred (prior) image~\ref{fig:TwoD_Blurred} and the noisy image~\ref{fig:TwoD_Noisy}, one can see that the posterior
    samples produce statistics those are closer to the original image.    
    Moreover, the retrieved results using the non-Gaussian \ClHMC (serial or parallel) algorithm is much better than the solution obtined by the traditional Tikhonov regularization approach.
    The relative error of the mean obtained using the parallel \ClHMC sampling algorithm is $0.01663$, while the relative error of the regularized solution is $0.021447$.
    The relative error is defined as 
    $
    \frac{\lVert \x -\x^{\rm true} \rVert}{\lVert \x^{\rm true} \rVert}\,,
    $
    where $\x$ is the retrieved image, and $\x^{\rm true}$ is the true image.
    The proposed parallel algorithms achieve an improvement of $29\%$ over the optimally tuned Tikhonov-based solution.
    These results show the capability and accuracy of the \ClHMC sampling algorithm in non-Gaussian settings.

    While we have shown the blurred image in Figure~\ref{fig:TwoD_Inputs}, it is of utmost importance to highlight the fact that the formulation and the sampler is unaware of this blurred image.
    This is mainly due to the fact that we have uniformly sampled the Gaussian sample centered around this blurred image.
    To further enhance the retrieved image, one can follow the inverse problem solution by deblurring filter which is out of the scope of this work.
	
	
\section{Conclusions and Future Work} \label{Sec:Conclusions}		
  %
  In this work, we have developed parallel cluster sampling algorithms for solving Bayesian inverse problems.
  Specifically, we have proposed parallel sampling algorithms based on the cluster sampling filter (\ClHMC) developed mainly for non-Gaussian data assimilation.
  The proposed algorithms can be efficiently used for solving various large-scale problems including medical image retrieval from noisy observations.
  
  %
  We have introduced a detailed complexity analysis of the proposed parallel clustering sampling (\ClHMC) algorithms with mixture model represenation of the prior information.
  Generally speaking, aside from parallelization, the parallel versions of the algorithm result in higher acceptance rates.
  Specifically, the parallel \ClHMC  increases the acceptance rate of the sampler from $44\%$ to $93\%$ with Gaussian proposal kernel, leading to massive saving of computations.
  The proposed sampling algorithms achieve an improvement of $29\%$ over the optimally-tuned Tikhonov-based solution for image retrieval.
  The algorithm can run significantly faster than the serial sampler in ideal settings.
  However, the algorithm can be slower than the serial sampler if too many outliers exist where some chains are assigned much smaller ensemble size than the others.
  The \ClHMC sampling algorithm, in addition to desired parallelization features, has proved powerful in the context of Bayesian image retrieval.
  
  In future work, we will investigate the possibility of parallelizing  other components of the sampling algorithm such as the likelihood function and the proposal mechanisms.
  For example a parallel version of EM can be considered to construct the GMM approximation to the prior distribution. 
  In the case of HMC, the symplectic integrator can be parallelized. 
  Also, matrix-vector products can be parallelized. 
  Methods for parallelizing a single chain can be considered for a second level of parallelization. 
  While our implementation is not a direct parallelization of the MCMC algorithm, it still provides acceptable sampling with better 
  performance. 
  

\bibliographystyle{plain}
\bibliography{Bib/data_assim_HMC,Bib/General,Bib/Software,Bib/data_assim_kalman}


\end{document}